\documentclass[3p,times]{elsarticle}

\usepackage{siunitx} 
\DeclareSIUnit\mK{\milli\K}
\DeclareSIUnit\ueV{\micro\eV}
\DeclareSIUnit\micron{\micro\meter}
\DeclareSIUnit\sec{\second}
\DeclareSIUnit\parsec{pc}
\DeclareSIUnit\mpc{\mega\parsec}
\DeclareSIUnit\kpc{\kilo\parsec}
\DeclareSIUnit\torr{torr}
\DeclareSIUnit\psi{psi}
\DeclareSIUnit\inch{inch}
\DeclareSIUnit\sccm{sccm}
\DeclareSIUnit\rpm{rpm}
\DeclareSIUnit\oz{oz}
\DeclareSIUnit\n{n}
\DeclareSIUnit\year{yr}
\DeclareSIUnit\each{ea}
\DeclareSIUnit\bq{Bq}
\DeclareSIUnit\gauss{G}
\DeclareSIUnit\gcc{\gram\per\cubic\cm}
\DeclareSIUnit\uW{\micro \watt}
\DeclareSIUnit\nW{\nano \watt}
\DeclareSIUnit\kOhm{\kilo \ohm}
\usepackage{pgfplots}
\pgfplotsset{compat=newest}
\usetikzlibrary{plotmarks}
\usetikzlibrary{arrows.meta}
\usepgfplotslibrary{patchplots}
\usepackage{grffile}
\usepackage{amsmath}
\usepackage{fix-cm}
\usepackage{relsize}

\usepackage{ctable}
\usepackage{multirow}
\usepackage{textcomp} 

\usepackage{amssymb}

\biboptions{square}

\setupctable{captionsleft} 

\usepackage{enumitem}
\usepackage{xfrac}

\usepackage[printwatermark]{xwatermark}
\usepackage{xcolor}
\usepackage{graphicx}

\usepackage{soul} 
\setul{1pt}{.4pt} 
\usepackage[utf8]{inputenc} 

\begin{document}

\begin{frontmatter}



\title{Properties of selected structural and flat flexible cabling materials for low temperature applications}
\author[ucsb]{M. Daal\corref{cor1}}
\author[ucsb]{N. Zobrist}
\author[colorado1,colorado2]{N. Kellaris}
\author[ucb]{B. Sadoulet}
\author[ucd]{M. Robertson}
\address[ucsb]{Department of Physics, University of California, Santa Barbara CA 93106-9530}
\address[ucb]{Department of Physics, University of California, Berkeley CA 94720-7300}
\address[ucd]{Department of Physics, University of California, Davis CA 95616-8677}
\address[colorado1]{Department of Mechanical Engineering, University of Colorado Boulder, 80309-0427}
\address[colorado2]{Materials Science and Engineering Program, University of Colorado, Boulder, 80309-0596}
\cortext[cor1]{mdaal@berkeley.edu}
\begin{abstract}

We present measurements of the low temperature thermal conductivity for materials useful in the construction of cryogenic supports for scientific instrumentation and in the fabrication of flat flexible cryogenic cabling. The materials we measure have relatively low thermal conductivity. We present a method for measuring the heat transfer coefficient of flat cabling and show, using an example, that the thermal conductivity of a flex cable is reasonably well predicted by composing the thermal conductivities of its constituent material layers.  Room temperature physical and mechanical data is given for the materials studied, as well as an overview of relevant materials science and manufacturing details. Materials include Timet Ti 15-3 and Ti 21S, Materion alloy vit105 (LM105) in amorphous state, ATI Metals Nb-47Ti, Johnson Matthey nitinol (NiTi), Mersen graphite grade 2020, DuPont Pyralux coverlay and Vespel SCP-5050, and Fralock Cirlex polyimide sheets. All data is in the temperature range \SIrange[range-phrase = { to }]{0.05}{2}{\K}, and up to \SI{5}{\K} for SCP-5050.
\end{abstract}

\end{frontmatter}

\section{Introduction}\label{sec:intro}
\begin{figure*}[ht]
	\centering
  	\includegraphics[width=\textwidth]{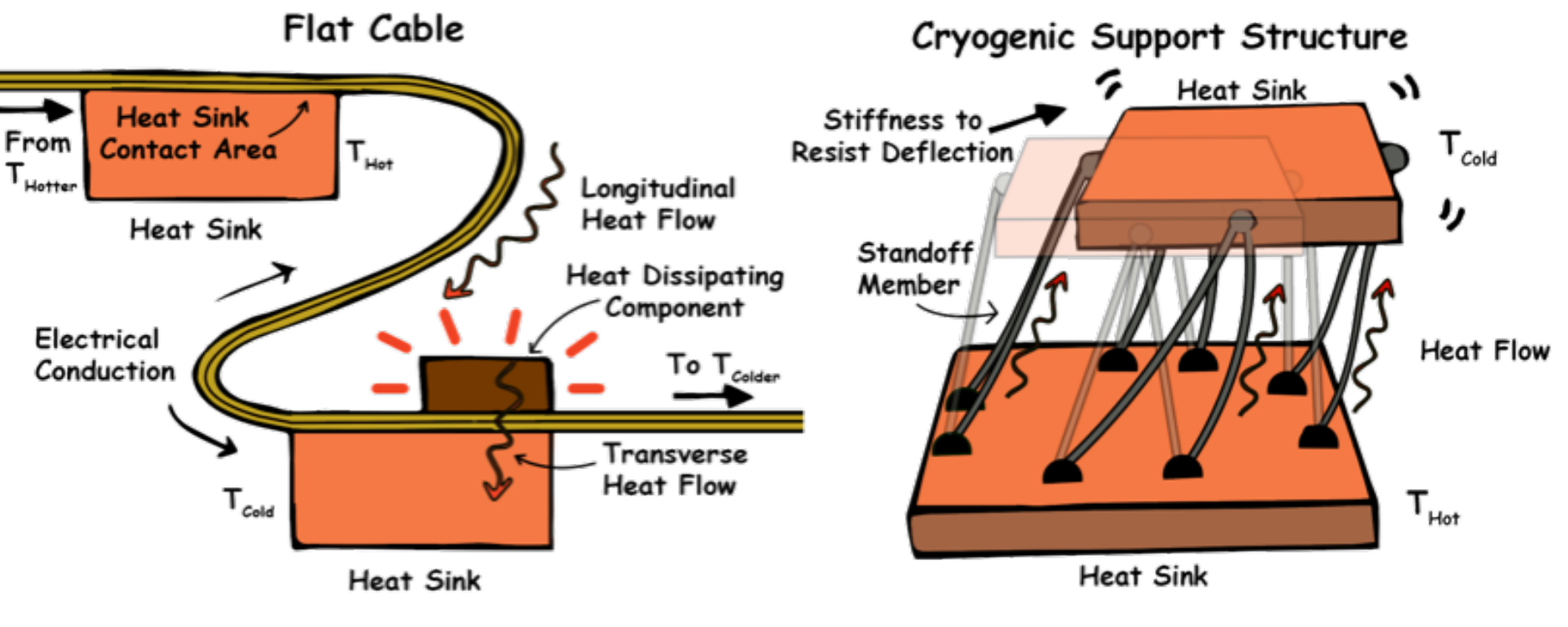}
  	\caption{ Left: A schematic drawing of a flat cable connecting two cryostat temperature stages, T$_\text{Hot}$ and T$_\text{Cold}$. \emph{Longitudinal heat flow} occurs along the cable from T$_\text{Hot}$ to T$_\text{Cold}$. The cable's longitudinal thermal conductivity quantifies this heat flow.  \emph{Transverse heat flow} occurs from one face of the cable to the other face of the cable, due to a heat dissipating component.  Right: A schematic drawing of an example structure providing stiff support between two cryostat temperature stages, T$_\text{Hot}$ and T$_\text{Cold}$. Heat flows through standoff members.}
  	\label{Fig:cable_and_support_structure}
\end{figure*}

This work is motivated by the need to design a cryogenic support structure called the `tower'  and low inductance, high trace count ($\num{\sim110}$), cabling for the detectors of the SuperCDMS (Cryogenic Dark Matter Search) SNOLAB experiment,~\cite{Agnese2017}. The tower provides heat sink attachment points as well as mechanical support for the detectors and their associated cabling. It is connected to each temperature stage of the cryostat, and must provide stiff mechanical support between stages while minimizing heat flow between them and being composed of very low radioactivity materials, Figure~\ref{Fig:cable_and_support_structure} (Right). The towers and cryostat are designed to maintain the detectors at \SI{0.015}{\K}. In order to complete the design, it was necessary to measure the sub-kelvin thermal conductivities of many candidate materials for possible use in the detector signaling cable and tower. 

Flat flexible cabling, made by laminating patterned metallic foil between insulating polymer layers, is chosen to convey detector signals because it allows one to have large trace counts with engineered impedances within a small cross section. For such cables, one wants to know the cable's thermal conductivity due to longitudinal heat flow, as well as its heat conductivity due to transverse heat flow, which can be quantified as a heat transfer coefficient, Figure~\ref{Fig:cable_and_support_structure} (Left). We discuss usage of the heat transfer coefficient in section~\ref{sec:boundary_res}. In designing these cables, the challenge is to minimize longitudinal heat flow while satisfying one's electrical impedance  and shape (e.g. length) requirements. To do this, one would ideally like to use the thermal conductivity of the constituent material layers to predict the aggregate longitudinal heat flow.    

We apply the same measurement techniques to different specimen geometries to measure the thermal conductivity of materials useful for flat flexible cabling and cryogenic support structures. These techniques are also used to measure the heat transfer coefficient of a multilayer laminated specimen due to transverse heat flow. In pursuit of minimizing the heat flow between cryostat temperature stages, the materials we selected to measure were expected to have low thermal conductivity, a priori. Error is estimated for all these measurements.

This article provides both low temperature heat conductivity data and higher (e.g. room) temperature material characteristics for a few materials of potential usefulness in constructing a low temperature instrument.  The article is organized as follows: In section~\ref{sec:materials} we present extensive background information on the materials measured including notes on sample preparation, how they are manufactured and the materials science regarding their room and low temperature properties. Readers principally interested in the thermal conductivity results may want to skip this section or only read the parts of it that pertain to the materials of interest to them.  In section~\ref{sec:procedure}, we discuss the details of our experimental setup and measurement. In section~\ref{sec:results}, we present our measurement results. In section~\ref{sec:conclusion}, we summarize our conclusions.



\section{Materials tested}\label{sec:materials}
The materials tested can be divided into those which are composed of thin component materials which have been laminated together, the `laminated materials', and those which are themselves `component materials' in either thin sheet or bulk form. Unless otherwise noted, all alloy compositions are given as weight percentages, i.e. X-nY-mZ is (100-n-m wt\%) X -- (n wt\%) Y -- (m wt\%) Z.

Often, cryogenic cabling and supports are expected to perform differently at room  temperature than they do at low temperature. For example, strength may be most important for a cryogenic support structure when it is at room temperature and being handled, while stiffness and thermal conductivity may be most important at low temperature~\cite{Runyan2008a, Kramer2014, Hastings1993}, when bolted in the cryostat. For this reason, we tabulate mechanical and physical properties of the component materials studied near room temperature in Table~\ref{Tab:Mechanical_Props}. Next,  we review background information on the materials science, manufacturing and the use at low temperatures of the materials we measured so as to aid the designer in making the most appropriate material selection for their application.

\subsection{Component materials}\label{sec:component_materials}
\sisetup{
range-phrase = --
}
\ctable[
		cap = Mechanical properties of selected materials,
		caption = {Yield strength (YS) [at \SI{0.2}{\%} offset strain unless otherwise noted], ultimate tensile strength (UTS), ultimate elongation (UE), tensile modulus (TM), density ($\rho$), coefficient of thermal expansion (CTE), hardness (H) [Vickers H$_{\text{v}}$; Rockwell B, E, C and L; Shore scleroscope SSH], maximum continuous operating temperature in standard pressure air (T$_{\text{max}}$), glass transition temperature (T$_{\text{g}}$), and superconducting transition temperature (T$_{\text{c}}$) are tabulated for the materials whose thermal conductivity is presented in this work. Test methods are indicated when available. `Ultimate' generally refers to measurements at material failure; whether this is a yield or break is indicated in some cases for clarity.  Typical (average) values for YS, UTS, UE, TM and $\rho$ are at standard temperature and pressure unless otherwise noted.  Conditions for these properties vary and are called out below. `---' indicates that value is unavailable or not applicable; `$\uparrow$'/`$\downarrow$' indicates that value is unavailable but most likely similar to above/below value. T$_\text{C}$ measurements are described in section \ref{sec:Tc}; on pieces of the materials described in section~\ref{sec:component_materials}

\ul{\emph{Pyralux LF.}}  UTS, UE and TM for Pyralux, YS for Kapton HN. $K_{\SI{300}{\K}}$ for Kapton HN and adhesive, respectively.  T$_{\text{max}}$ is for Pyralux FR (`Flame retardant') which has a specification on this quantity as part of its UL flammability (VTM-0) rating. LF is not rated, but these adhesives are similar enough to justify using the FR value for LF. CTE is for \SIrange[range-phrase = { to }]{0}{40}{\degreeCelsius}. H is pencil hardness to scratch 100HN film. Electrical: at \SI{1}{\MHz} $\epsilon_\text{r}$ \num{3.8}\tmark[$\dagger$] and $\tan \delta$ \num{0.030}\tmark[$\dagger$]. All data from \cite{DataSheet}.

\ul{\emph{Cirlex.}}   T$_{\text{max}}$  = T$_{\text{g}}$, but \SI{400}{\degreeCelsius} is permissible for short durations. Electrical: at \SI{10}{\kHz} $\epsilon_\text{r} =$ \num{3.45} and $\tan \delta =$ \num{0.0021}. All data from \cite{DataSheet}. $\perp$/$\parallel$ is with respect to the plane of the sheet.

\ul{\emph{SCP-5050.}} CTE is from \SI{23}{\degreeCelsius} to \SI{300}{\degreeCelsius}. $K_{\SI{300}{\K}}$ is at \SI{50}{\degreeCelsius}. T$_{\text{max}}$ temperature at which there is a 50\% reduction in tensile strength after \SI{10000}{\hour} exposure in air.  All data from \cite{DataSheet}.

\ul{\emph{2020 graphite.}} CTE is for \SIrange[range-phrase = { to }]{400}{500}{\degreeCelsius}. Additionally, shear modulus and poisson ratio are \SI{4.7}{\GPa} and \num{0.17}, respectively. UE from \cite[][pg. 3-19]{Ho1988}, all other data from \cite{DataSheet}

\ul{\emph{LM105.}} CTE is for \SIrange[range-phrase = { to }]{577}{1027}{\degreeCelsius}. UTS, YS dependent on post-cast surface preparation (e.g. removal of flash) and are equal because there is little or no yielding before failure. Poisson ratio is \num{0.38}. All data from \cite[][e.g.liquidmetal.com as of 07/2018]{DataSheet}.

\ul{\emph{Nb-47Ti.}} Nb-47Ti = 36.76 at\% Nb--63.24 at\% Ti. YS, UTS, UE, H from \cite[Tab. 3 for 46.5 wt\% Ti]{Curtis1979} for annealed condition. (YS, UTS can almost double with  cold work.)  TM estimated from \cite[Tab. 3.4, for 64 at\% Ti]{Collings1986}. CTE estimated from \cite[][pg. 51]{Clark1975} for Nb-48Ti \SIrange[range-phrase = { to }]{17}{27}{\degreeCelsius}. $K_{\SI{300}{\K}}$  from \cite{Flachbart1977a} for Nb-50Ti.

\ul{\emph{Ti 15-3.}} All values for strip stock; CTE and $K_{\SI{300}{\K}}$ may be slightly anisotropic due to rolling. We assume solution treatment at \SIrange[range-phrase = { to }]{785}{840}{\degreeCelsius} for \SIrange{4}{30}{\minute} then air cool: H from \cite{Boyer1998}; T$_{\text{max}}$ from \cite{Nyakana2005}; T$_{\text{c}}$ from solution treated sheet sample. All other properties from \cite{Fanning1993b}; CTE is for  \SIrange[range-phrase = { to }]{23}{100}{\degreeCelsius}. We assume aging at \SI{482}{\degreeCelsius} for \SI{16}{\hour}; H from \cite{Boyer1998}; T$_{\text{max}}$ from \cite{Nyakana2005}; all other properties from \cite{Fanning1993b}.

\ul{\emph{Ti 21S.}}  All values for strip stock. YS, UTS, UE, TM from \cite[][e.g from timet.com as of 07/2018]{DataSheet}.  We assume solution treatment at \SIrange[range-phrase = { to }]{816}{843}{\degreeCelsius} for \SIrange{3}{30}{\minute} then air cool: T$_{\text{max}}$ and CTE for interval \SIrange[range-phrase = { to }]{25}{38}{\degreeCelsius} are from \cite{Nyakana2005}.  We assume aging at \SI{538}{\degreeCelsius} for \SI{8}{\hour}: $\rho$, $K_{\SI{300}{\K}}$ from \cite{Nyakana2005}. 

\ul{\emph{NiTi.}} General values for material annealed at unspecified time and temperature of the 55Ni-Ti alloy family. Values for cold worked or aged material vary substantially. YS, UTS, TM, $K_{\SI{300}{\K}}$ from \cite{ASMInternational-SMA}. UE from \cite{Jackson1972}. H, $\rho$, CTE and poisson ratio is \num{0.33} from \cite[][pp. 1035-1048]{Boyer1998}. Austenite values assumed to be at room temperature, which is less than M$_\text{d}$, and martensite values assumed to be below room temperature. Broad superconducting transition occurred over the range \SIrange{0.217}{0.237}{\K}.

\ul{\emph{SS 316.}} Included for comparison. YS, UTS, UE and H for annealed sheet specimen;  UE is on \SI{50.8}{\mm} [\SI{2}{\inch}]  sample; CTE is from \SIrange[range-phrase = { to }]{0}{100}{\degreeCelsius}; $K_{\SI{300}{\K}}$ is at \SI{100}{\degreeCelsius}; data from \cite{Harvey1982} except T$_{\text{max}}$ from \cite[][pg.511]{Davis1994}. Type 316 is a molybdenum containing austenitic chromium-nickel stainless steel offering superior corrosion and high temperature resistance in comparison to other 300 series steels. Type 316LN (Figure~\ref{Fig:Support_Mat_K}) has lower carbon and higher nitrogen content than type 316, but the two share the same minimum YS, UTS, UE and H specifications per ASTM A240. The extra nitrogen is linked to an increase in YS and UTS at \SI{4.2}{\K} with only moderate decrease in UE \cite{Sas2015}. 

\ul{\emph{Cu 10100}} Included for comparison. Oxygen-free electronic copper, OFE (formerly OFHC), is \SI{99.99}{\%} Cu minimum.  Values for flat mill product, cold worked. YS is at 0.5\% offset; T$_{\text{max}}$ is where YS and UTS are about half their \SI{20}{\degreeCelsius} values; UE is on \SI{50.8}{\mm} [\SI{2}{\inch}]  sample; CTE is from \SIrange[range-phrase = { to }]{20}{100}{\degreeCelsius}. All values from \cite{Robinson1990}.

		},
        label   = Tab:Mechanical_Props,
		topcap,
		sideways,
		doinside = \footnotesize, 
		notespar 
]{lrrrrrrrrrrrr}{
 \tnote[a]{3\% offset strain}
 \tnote[b]{ASTM D882}
 \tnote[c]{ASTM F443}
 \tnote[d]{ASTM D638}
 \tnote[e]{ASTM D1505}
 \tnote[f]{ASTM D831}
 \tnote[g]{ASTM D729. $\rho$ is obtained from specific gravity using the density of water, \SI{0.997}{\gcc}, at standard temperature and pressure.}
 \tnote[$\dagger$]{IPC 2.5.5.3}
 \tnote[B]{Ultimate refers to break}
 \tnote[$\ddagger$]{Absolute thermometer uncertainty is \SI{\pm 0.007}{\K} for Nb-47Ti, \SI{\pm 0.02}{\K} for NiTi and \SI{\pm 0.05}{\K} for the other samples.}
}{
\FL
             &YS         &UTS        &UE        &TM        &$\rho$     &CTE                 &H        &T$_{\text{max}}$ &T$_{\text{g}}$ &T$_\text{C}$\tmark[$\ddagger$] &$K_{\SI{300}{\K}}$     & Material \NN
             &[\si{MPa}] &[\si{MPa}] &[\si{\%}] &[\si{GPa}]&[\si{\gcc}]&[\SI[per-mode=repeated-symbol]{e-6}{\per \degreeCelsius}]&          &[\si{\degreeCelsius}] &[\si{\degreeCelsius}] &[\si{\K}] &[\si{\watt \per \meter \per \K}]&Condition         \ML

Pyralux LF\tmark[B]   &\num{69}\tmark[a]\tmark[b]  &\SIrange[range-units = repeat]{69}{138}{\tmark[b]}&\num{70}\tmark[b] &\SIrange[range-units = repeat]{1.03}{2.07}{\tmark[b]} &\SIrange[range-units = repeat]{1.3}{1.4}{\tmark[e]} &\num{100} &\num{2}B&\num{105}&\num{40}&\num{}--- &\num{0.12}\tmark[c], \num{0.23}& ---\NN 

Cirlex      &\num{49}\tmark[a]   &\num{230}\tmark[d]  &\num{57}\tmark[d]  &\num{1.9}\tmark[d]    &\num{1.42}\tmark[e] &\num{30}$^\parallel$\tmark[f], \num{118}$^\perp$\tmark[f] &\num{}--- &\num{351} &\num{351} &\num{}--- &\num{0.17} & ---\NN
SCP-5050\tmark[B]     &\num{}---&\num{72}\tmark[d] &\num{2.5}\tmark[d] &\num{8.9}\tmark[d] &\num{1.75}\tmark[g] &\num{29}\tmark[f]   &\num{63}E   &\num{288} &\num{320} &\num{}---    &\num{1.44}\tmark[c]& Iso. Molded \NN 
2020 graphite\tmark[B]&\num{29}  &\num{}---    &\num{\approx 0.3}    &\num{11} &\num{1.77} &\num{4.3}            &\num{95}L,\num{52}SSH     &\num{350} &\num{}--- &\num{}---   &\num{85} & Iso. Molded\NN 
LM105\tmark[B] &\numrange{1524}{1850} &\numrange{1524}{1850}&\num{1.8} &\num{92.7}  &\num{6.68} &\num{12.0}         &\num{563}H$_{\text{V}}$, \num{53}C  &\num{250} &\num{399} &\num{1.25} &\num{}--- & Amorphous  \NN
Nb-47Ti    &\num{446}&\num{464}&\num{25}&\num{82}&\num{6.02}&\num{9}     &\num{139}H$_{\text{V}}$&--- &\num{}--- &\num{8.851} &\num{10} & ---\NN 
Ti 15-3      &\num{789}&\num{791}&\num{14.4}&\num{82}  &\num{4.71} &\num{8.4}           &\num{\approx290}H$_{\text{V}}$&\num{288} &\num{}--- &\num{3.65} &\num{8.08}  & Sol. Treated  \NN 
             &\num{1296}&\num{1393}&\num{6.3}&\num{109}  &\num{}$\uparrow$ &\num{}$\uparrow$          &\num{\approx340}H$_{\text{V}}$&\num{}$\uparrow$ &\num{}---&\num{}$\uparrow$ &\num{}$\uparrow$ & Aged \NN 
Ti 21S       &\num{883}&\num{931}&\num{12}&\numrange{72}{85}  &$\downarrow$ &\num{}$\downarrow$          &\num{}---&\num{593} &\num{}--- &\num{3.10} &\num{}$\downarrow$ & Sol. Treated  \NN 
             &\num{1338}&\num{1427}&\num{6}&\numrange{103}{110}  &\num{4.93}&\num{7.07}           &\num{}---&\num{}$\uparrow$ &\num{}--- &\num{}$\uparrow$ &\num{7.6} & Aged \NN 
NiTi      &\numrange{195}{690}&\num{895}&\numrange{10}{60}&\num{83}&\num{6.45}&\num{11.0}     &\num{190}H$_{\text{V}}$ &\num{}--- &\num{}--- &\num{}---&\num{18} & Austenite\NN 
          &\numrange{70}{140} &\num{}---  &\num{}--- &\numrange{28}{41}&\num{6.46}&\num{6.6}      &\num{190}H$_{\text{V}}$ &\num{}--- &\num{}--- &\num{0.227} &\num{8.5} & Martensite\NN
SS 316     &\num{290} &\num{580} &\num{50} &\num{193} &\num{8.0} &\num{16.0}             &\num{79}B   &\num{925}   &\num{}---&\num{}--- &\num{16.3} & Annealed\NN 
Cu 10100     &\num{205} &\num{260} &\num{25} &\num{115} &\num{8.94} &\num{17.0}             &\num{25}B   &\num{360}   &\num{}---&\num{}--- &\num{391} & H01, $\sfrac{1}{4}$ Hard \LL
}
\sisetup{
range-phrase = { to }
}

\ul{\emph{Pyralux LF:}}
We present data on DuPont Pyralux LF coverlay composite, which consists of DuPont Kapton type HN film coated on one side with a proprietary temperature activated, `B-stage', acrylic adhesive. Kapton is the trade name for DuPont's family of polyimide film products. The primary use for the Kapton and adhesive composite films is in the fabrication of flexible circuits where the Kapton is used as the flexible dielectric substrate. In the terminology of the industry, a coverlay is a composite film that has a layer of dielectric and a layer of adhesive. A coverlay is used to encapsulate etched layers of flexible and `rigid flexible' (i.e. has stiffened portions) multilayer constructions for environmental and electrical insulation. A bondply is a layer of adhesive, then a layer of dielectric, then another layer of adhesive. Temperature activated adhesive films with no dielectric layer are also available. All of these products are available at various Kapton and adhesive thicknesses.

Our test specimen, Pyralux LF coverlay product code 0110, is \SI{25.4}{\micron} [\SI{0.001}{\inch}]  thick adhesive on \SI{25.4}{\micron} thick Kapton. To make the test sample, two sheets of LF 0110 were laminated together such that the acrylic adhesive of one sheet was adhered to the acrylic adhesive of the other (i.e. the stack-up is \SI{25.4}{\micron} Kapton, then \SI{25.4}{\micron} adhesive, then \SI{25.4}{\micron}  adhesive, then \SI{25.4}{\micron} Kapton).  As recommended by DuPont, the lamination requires heating the part to \SIrange{182}{199}{\degreeCelsius} and applying \SIrange{200}{400}{\psi} pressure for \SIrange{1}{2}{\hour}. The laminated construction was folded upon itself and  heat was directed parallel to the surface of the sheets as done in \cite{FilmSampleGeo}. Although Kapton HN  films are known to be semicrystalline with the molecular axes aligned in the plane of the film and close to its `machine direction' (probably within \SI{\approx 30}{\degree} of the machine direction~\cite{DataSheet}, which is the long direction in which the material was produced and rolled), no attempt was made to keep track of the machine direction with respect to the direction of heat flow.

Pyralux composites meet the criteria of NASA outgassing specification, SP-R-0022, with \SI{\le 1.0}{\%} total mass loss (TML) and \SI{< 0.10}{\%} collected volatile condensable material (CVCM), \cite{NASAOutGassingDB}.

\ul{\emph{Cirlex:}}
The specimen is from a \SI{0.787}{\mm} [\SI{0.031}{\inch}] thick sheet provided by Fralock -- currently DuPont's sole licensee for the production of Cirlex worldwide. Our interest in Cirlex stemmed from its low radioactivity, relative to other circuit board laminates. Heat flow was parallel to the plane of the  sheet for the thermal conductivity measurement.  Cirlex is polyimide material currently available in panels \SIrange{0.102}{4.724}{\mm} [\SIrange{0.004}{0.186}{\inch}] thick. Panels are available with or without copper cladding for use in printed circuit board manufacturing. The material has mechanical properties  (Table~\ref{Tab:Mechanical_Props}) suitable for use in structural elements and as stiffeners in flexible circuit applications.  DuPont Kapton HN and Cirlex are made from the same polyimide monomer precursor film except that the Cirlex precursor film is treated with organotin stabilizer added at few parts per million concentration, whereas Kapton HN is cured precursor. Fralock stacks together the precursor films to obtain different product thicknesses. Aggressive heat and pressure, not adhesive, are used to bond the films together and to drive out volatiles from the precursors. Considering this manufacturing process, it is not surprising that Cirlex and Kapton HN have similar mechanical, physical, thermal, chemical and electrical properties.

Cirlex also meets SP-R-0022, with \SI{1.134}{\%} TML, \SI{< 0.013}{\%} CVCM and \SI{0.753}{\%} water vapor regain according to \cite[][on a \SI{3.175}{\mm}, \SI{0.125}{\inch}, thick specimen]{DataSheet}. The precursor films and the finished product are semicrystalline.

\ul{\emph{Vespel SCP-5050:}}
DuPont Vespel is a family of high performance plastics which have a polyimide base polymer. The Vespel SCP line was introduced in the early 2000s as an improvement to the Vespel SP line which was commercialized in the 1960s. In comparison to other performance plastics and to the SP products, Vespel SCP products are attractive for their sustained mechanical and electrical stability at high temperatures, generally \SI{\ge 250}{\degreeCelsius}, and for long durations; their chemical resistance, although they are susceptible to alkaline attack; their radiation resistance; low outgassing in vacuum environments, following an initial bake out (SCP-5050 meets SP-R-0022 with \SI{0.13}{\%} TML and \SI{0.01}{\%} CVCM and \SI{0.03}{\%} water vapor regain \cite{NASAOutGassingDB}); and their low coefficient of thermal expansion at elevated and cryogenic temperatures. Generally, the SCP products outperform the SP in all these areas, often considerably. On the other hand, Vespel SCP material is relatively expensive. While Vespel SP and SCP materials do not melt, the SCP line has a glass transition temperature at approximately \SI{320}{\degreeCelsius}.

Vespel products are available with certain filler materials added to  achieve different properties.  SP-1 and SCP-5000 are unfilled polyimide polymers. Graphite is added to the polyimide base material in grades SP-21 (15 wt\%), SP-22 (40 wt\%), SCP-5050 (unspecified wt\%). The graphite filler results in lowered cryogenic thermal conductivity~\cite{Kellaris2014, Runyan2008a}, coefficients of thermal expansion and friction, and improved wear resistance.

Vespel can be purchased in several stock shapes including rod, tube, bar and plaque. Upon request, DuPont can also produce custom Vespel shapes and parts. DuPont uses different manufacturing processes to make stock and custom shapes and parts. One of these processes is hot isostatic molding where heat and pressure are applied from all sides of powder in a mold until it attains a desired form. `Isostatic' parts and shapes, made by this process, are isotropic in their properties, e.g. thermal conductivity and coefficient of thermal expansion (CTE). Rod, tube and bar are manufactured via hot isostatic molding.  Another process used is standard compression molding, where only pressure is applied to the powder and then a post-forming sintering operation is performed to strengthen the material. `Direct form' parts and shapes, made by this process are (sometimes considerably) anisotropic in their properties.

Thermal conductivity data was obtained from stock rod material made by the isostatic process, which was originally manufactured by DuPont and distributed by Curbell plastics. Since there are many imitation products on the market, it is advisable to request certified `authentic Vespel' through an authorized DuPont distributor when sourcing this material.

\ul{\emph{2020 Graphite:}}
The sample measured was manufactured by and obtained from Mersen USA corporation. The manufacturing process of grade 2020 involves forming stock shapes via cold isostatic pressing of particle grains\footnote{In the terminology of the graphite industry, `grain', refers to a particle of filler material, \cite{ASTMC709}.} of ground filler material (which may come from petroleum cokes, natural and/or recycled graphite) mixed with a pitch binder~\cite{Burchell2007}, baking and then graphitization at \SI{>3000}{\degreeCelsius}.  The filler material average particle grain size is \SI{\approx 15}{\micro\meter}, and the finished graphite is porous with \SI{9}{\%} of its volume composed of voids. This grade is marketed as an isotropic artificial graphite for use in the pressure sintering industry as a mold material, and as a die material in hot pressing applications --- applications where its excellent machinability and dimensional stability at high heat are exploited.

The forming process imparts a preferred particle grain orientation into the finished graphite as the long axes of filler particles align perpendicular to the direction of the forming force. The long axis of the particles is usually parallel to the graphene planes of its graphite crystal structure (i.e. aligned to a-axis), however the particle eccentricity depends on the source of the particular filler material used and can be minimized by selection,~\cite{Ho1974}. In the end, there will be a direction in the graphite which is maximally aligned with the c-axis (the `across grain' direction in the parlance of the industry, per ASTM C709, and denoted $\perp$ here) and a direction which is maximally aligned with the a-axis (the `with grain' direction and denoted $\parallel$ here).

The samples we measured came from a \SI{101.6 x 101.6 x 101.6}{\mm} [\SI{4 x 4 x 4}{\inch}]  cube section of billet product, which was marked by Mersen to indicate its `with' and `across grain' directions. The cube was halogen purified by Mersen to a spot total purity of \SI{\approx 1.3}{ppm} measured elemental contaminants by weight. Two rods were cut from the cube, one with its axis oriented with the grain and the other with its axis oriented across the grain. The thermal conductivity of each rod was measured with heat flowing parallel to the rod axis. The thermal conductivity data for heat flowing across the grain as well as a description of the purification process, is presented in Kellaris et al~\cite{Kellaris2014}. We present thermal conductivity data for heat flowing in the with grain direction.

\ul{\emph{65.7Zr-15.6Cu-11.8Ni-3.7Al-3.3Ti (Common Name LM105, vit105):}} The Materion Corporation produces this alloy in crystalline ingot form. Liquidmetal Technologies uses the ingot as feedstock in an injection molding process to produce an amorphous alloy which we used for our measurements\footnote{The Materion and Liquidmetal designations for this alloy are vit105 and LM105, respectively}. This alloy is one of a family of alloys which can be cooled from a liquid at a relatively slow rate without the sudden formation of a crystalline phase and therefore yields a frozen liquid atomic configuration throughout the bulk of the sample ~\cite{Wang2004,Chen2011}, i.e. a bulk metallic glass. Consequently, it can be molded into complex geometries, including small cavities, with minimal post-processing, minimal shrinkage (since there is no crystal phase transformation), and excellent repeatability. LM105 is an example of a commercially available Zr-based metallic glass; other metallic  glasses such as Cu-, Pd-, Ni-, Fe-  as well as lower density~\cite{Jiang2015} Al-, Mg-, Ti-based  materials exist but may not be as commercially available.  Generally, these materials have high yield strength, elastic limits ($\sim$2\%), hardness, and corrosion resistance. The high elastic limit is concomitant with a high elastic storage capability, with minimal loss. Bulk metallic glasses exhibit an increase in plasticity (ductility or fracture toughness) as their temperature is lowered~\cite{Sun2015}, a potentially useful characteristic for cryogenic structural support applications. Though the amorphous phase is not thermodynamically stable, they exhibit metastability at and below room temperature for sufficiently long times to allow their use in structural applications.

The specimens measured are as-injection-molded trapezoidal prisms. The surfaces have been slightly lapped to ensure flatness and to remove any flash. Heat flow was directed down the long axis of the prisms. Based on previous samples formed by the same process and using the same mold, the sample measured is expected to be \SI{99}{\percent} amorphous by volume or greater.

\bigskip

\paragraph{Titanium Alloys}\mbox{}\\
Pure titanium has a hexagonal close-packed, `$\alpha$', crystal structure at room temperature, and a body-centered-cubic (bcc), `$\beta$', crystal  structure at temperatures above \SI{882.5}{\degreeCelsius}. The addition of certain alloying elements to titanium can change this temperature and allows for a mixed,  $\alpha+\beta$, phase to exist in equilibrium for temperatures below those where pure $\beta$ phase is in equilibrium. The temperature at which the pure $\beta$ phase decomposes to $\alpha+\beta$ phases is the `$\beta$ transus' (typically in the range \SIrange{750}{950}{\degreeCelsius} depending on the alloy). Alloying elements which decrease the $\beta$ transus are called $\beta$-stabilizers and examples include Mo, V, Nb, and Ni. Within the $\alpha+\beta$ temperature range,  $\alpha$ phase may be created by a nucleation-and-growth type mechanism which requires long range atomic movement, i.e. diffusion. Alternatively, $\alpha$ can be created as the result of cooperative short-range atomic rearrangements into a more stable crystal structure, i.e. a `martensitic' transformation. Depending on alloy composition, martensitic transformations either occur at lower temperatures than those where nucleation-and-growth type transformation occur or do not occur at all. In the former case, the martensite start temperature, M$_\text{s}$, is the temperature below which martensitic transformation occurs. In the latter case, the material  is called a metastable-$\beta$ titanium alloy, \cite{Collings1983,Collings1986,Boyer1998}. This is in contrast to yet another case where the addition of alloying elements has depressed  the $\beta$ transus below room temperature, and the  material is regarded as a (stable) $\beta$ titanium alloy. $\beta$ titanium alloys are attractive for their high strength-to-density ratios and their outstanding room temperature formability~\cite{Bania1993}. 
Metastable-$\beta$ alloys can, additionally, be heated to a temperature below the $\beta$ transus, typically between \SIrange{400}{600}{\degreeCelsius} for \SIrange{0.5}{16}{\hour}, to precipitate finely dispersed needle-like $\alpha$-phase particles (as well as other phases) within their $\beta$ matrix, \cite{Nyakana2005}. These heat treatments, known as `aging', have the effect of strengthening and hardening the material. The pure $\beta$ phase can be recovered by annealing the material above the $\beta$ transus and then rapidly cooling to prevent significant precipitation of $\alpha$ (typically air-cool or faster). Such heat treatment to remove precipitates is called `solution heat treatment' or `solution annealing'.  
Unfortunately, the bcc structure of $\beta$ alloys is generally subject to a loss of ductility, i.e. UE approaches zero, at low temperature~\cite[][sec. 1, part 8]{Boyer1998} in what is known as a `ductile-to-brittle transition'. The presence of interstitial contaminants, e.g. hydrogen, exacerbates the effect (\cite{Reytier2002} and \cite{Schutz1993}, \cite[][a good survey of low temperature pressure cell materials]{Walker2005}) and to be avoided. Rather than a sharp transition, a steady decline in plasticity is seen in $\beta$ alloys with extra low interstitial (`ELI' grade) content and commencing at lower temperatures than is the case of higher interstitial contaminated alloys.

\ul{\emph{53Nb-47Ti (Common Name Nb-47Ti):} }
The principle commercial application for Nb-47Ti is for use in superconducting wiring and cabling, such as those used in magnet wiring. With its $\beta$ transus in the range \SIrange{550}{600}{\degreeCelsius} \cite{Lee1999,Moffat1988a}, Nb-47Ti is suitable for this application because it is relatively inexpensive, has high ductility which facilitates wire drawing, and can be thermomechanically processed to support high critical current densities (up to \SI{3500}{\ampere \per \mm \squared} at \SI{5}{\tesla} and \SI{4.2}{\K},~\cite{Cardwell2003}) and critical fields at liquid helium temperatures. The high critical current densities are due to the ability of $\alpha$-titanium precipitates (non-superconducting at \SI{4.2}{\K}) to pin fluxoid motion~\cite{Meingast1989}. Conventionally processed Nb-47Ti wire strands are subject to a finely-tuned sequence of cold deformation and aging steps in order to obtain nanometer-scale $\alpha$-titanium precipitates whose spacing matches the fluxoid spacing for maximal pinning efficacy~\cite{Lee1999,Cardwell2003}. The specimens measured here were not processed in this way.

The choice of 47 wt\% titanium alloy allows magnet wire manufacturers enough titanium to attain a desired $\alpha$-titanium precipitate volume fraction (up to 25 vol\%, while depleting the titanium concentration in the surrounding $\beta$-Nb-Ti matrix to between \SIrange[range-phrase = { and }]{37}{38}{wt\%} \cite{Lee1999}), but not too much titanium as to enable the growth of unwanted metastable phases (e.g. $\omega$ phase) and precipitate morphology (e.g. Widmanstätten) during thermomechanical processing. The T$_\text{C}$ also changes as a consequence of conventional processing, \cite{Meingast1989a}. 
Based on the behavior of Nb-50Ti~\cite[][pg 142]{Collings1986}, as the temperature is lowered, Nb-47Ti may be expected to gradually lose ductility instead of a sharp brittle transition while exhibiting a concomitant two to three-fold increase in yield and tensile strengths.

The Nb-47Ti sample studied was alloyed, forged, and cold rolled to foil form by ATI Specialty Alloys \& Components\footnote{Contact custserv@ATImetals.com.} using the Nb-Ti impurity specification established for the superconducting supercollider, \cite[][2500  ppm-wt limit for the Ta; 46 to 48 wt\% Ti limit for 'Nb-47Ti']{ASM-IHC-NbTi}, though the actual ingot chemistries are typically much better for most elements.  It was last annealed at \SI{860}{\degreeCelsius} for \SI{20}{\minute}, and then cold rolled from \SI{254}{\micron} [\SI{0.01}{\inch}] thickness to its final thickness of \SI{63.5}{\micron} [\SI{0.0025}{\inch}]  on average. For the thermal conductivity measurement, heat flow was directed parallel to the surface of the foil as done in ~\cite{FilmSampleGeo} and perpendicular to the roll direction.


\ul{\emph{Ti-15V-3Cr-3Sn-3Al (Common Names Ti 15-3-3-3, Ti 15-3):}} Ti 15-3 is a metastable-$\beta$ titanium alloy with a $\beta$ transus in the range \SIrange{750}{770}{\degreeCelsius},  \cite{Nyakana2005, Boyer1998}. It was developed in the 1970s by the Titanium Metals Corporations, Timet, and Lockheed with support from the U.S. Air Force to be a weldable, age hardenable, room temperature formable sheet alloy such as to provide an economical (i.e. relatively low stock and fabrication costs) high specific strength strip material, \cite{Nyakana2005,Cotton2015,Bania1993,Fanning1993b}. The specimens studied come from material alloyed$^\text{\ref{foot:ELI}}$, milled to sheet form, and then solution treated by Timet. We present thermal conductivity data for \SI{1.016}{\mm} [\SI{0.040}{\inch}] thick solution treated and `pickled' (i.e. soaked in an acid solution) sheet mill product as well as a heat treated \SI{21.336}{\micron} [\SI{0.00084}{\inch}] thick foil which was rolled by Arnold Magnetic Technologies from \SI{0.635}{\mm} [\SI{0.025}{\inch}] thick Timet mill product to \SI{21.336}{\micron} finished thickness, which they annealed (\SI{899}{\degreeCelsius} for \SIrange{5}{15}{\minute}s) last at \SI{25.4}{\micron} [\SI{0.001}{\inch}] thickness.  An additional heat treatment\footnote{The foil was heated to \SI{400}{\degreeCelsius} in an argon atmosphere for \SI{10}{\minute} and then allowed to air-cool to room temperature, which took about \SI{250}{\minute}. This heat treatment is part of our recipe for electroplating copper onto the foil for use in flex circuit fabrication. We wanted to confirm that the treatment does not affect thermal conductivity and, according to our data in Figure~\ref{Fig:Cable_Mat_K}, it does not. The lack of effect on the thermal conductivity is consistent with the expectation of no microstructure transformation based on the time temperature transformation data presented in \cite[][Fig. 12]{Cotton2015} for Ti 15-3, and similarly in \cite[][Fig. 7]{Niwa1990}.} (which we denote h.t. in the data) at the \SI{21.336}{\micron} finished thickness was performed on the foil by collaborators.  For the thermal conductivity measurement, heat flow was directed parallel to the surface of the sheet sample as depicted in Figure~\ref{Fig:Sample} for the case of NiTi. Heat was also directed parallel to the surface  of the foil sample as done in \cite{FilmSampleGeo}. The cold rolling steps in the mill and foil processing of our samples are expected to impart anisotropy (e.g. grain texturing) in their material properties, including CTE and thermal conductivity. Our thermal conductivity measurements did not account for any such anisotropy.

\ul{\emph{Ti-15Mo-2.7Nb-3Al-0.2Si (Common Name Ti 21S):}} Ti 21S is a metastable-$\beta$ titanium alloy with a $\beta$ transus in the range \SIrange{793}{810}{\degreeCelsius}, \cite{Boyer1998}. It was designed to have better high temperature behavior than Ti 15-3. It has higher tensile strength and better creep resistance than Ti 15-3, and among the lowest CTE and highest oxidation resistance of the $\beta$ titanium alloys~\cite{Nyakana2005,Cotton2015, Bania1993, Fanning1993a}. It is also more expensive than Ti 15-3. Our interest in Ti 21S was initially prompted by its availability in thicker sheet mill product in comparison to Ti 15-3, an attractive characteristic for low temperature structural applications. The specimens studied come from material alloyed\footnote{\label{foot:ELI} The Timet contact acknowledged tells the authors that while Timet's 15-3 and 21S alloys are specified to have slightly greater interstitial maximum allowances than Ti 6-4 ELI (per ASTM B265, see \cite{Reytier2002} and \cite{Walker2005} for discussion of ELI), the typical interstitial concentrations for these alloys are well within the Ti 6-4 ELI maximums.}, milled to sheet form and then solution treated by Timet. The sample measured was \SI{2.286}{\mm} [\SI{0.090}{\inch}] thick sheet and heat flow was directed parallel to its surface. Again, any anisotropy imparted by mill processing was not tracked.

\ul{\emph{55.7Ni-43.89Ti (Common Name Nitinol, NiTi):}} The specimen studied is in sheet form and was obtained from Johnson Matthey Inc. To produce the sheet, Johnson Matthey rolled source material provided by a third party supplier to \SI{1.498}{\mm} [\SI{0.059}{\inch}] finished thickness, then performed a shape set anneal at approximately \SIrange{450}{550}{\degreeCelsius} to make the sheet flat, and then pickled the sheet.  They report an ultimate tensile strength of \SI{1365}{\mega \pascal} with \SI{11.2}{\percent} elongation at failure in the cold rolling direction at \SI{22}{\degreeCelsius}. For the thermal conductivity measurement, heat flow was directed parallel to the surface of the NiTi sheet, perpendicular to the roll direction.

This material is an intermetallic compound and belongs to a family of NiTi alloys with nearly equiatomic composition (usually with a few at\% excess of Ni) possessing superelastic (sometimes called pseudoelastic) and shape memory properties.  Whether it exhibits superelastic or shape memory behavior at room temperature depends sensitively on the exact ratio of Ni and Ti, the addition of dopants, and the thermomechanical treatment history of the individual specimen. These conditions determine transition temperatures between a mechanically weaker martensitic phase at low temperatures and a stronger austenitic phase at higher temperatures\footnote{For NiTi the austenite phase has a B2 structure up to about \SI{600}{\degreeCelsius}  whereupon it becomes bcc~\cite{Jackson1972}. Martensite is B19 prime.}, which can be within the approximate range of \SIrange{-100}{100}{\degreeCelsius}. For example, aging at \SIrange{250}{500}{\degreeCelsius} has the effect of producing Ni-rich precipitates, enriching the surrounding matrix with Ti, and increasing phase transition temperatures~\cite{Pelton2003}; solution anneals, typically at \SIrange{500}{600}{\degreeCelsius}, rapidly re-solutionize these precipitates, decreasing transition temperatures. As such, heat treatment can be used to tune transition temperatures. Additionally, the transition is hysteretic: transformation from martensite to austenite does not occur over the same temperature range as the transition from austenite to martensite; typical hysteresis for NiTi alloys ranges from \SIrange{20}{40}{\degreeCelsius}.

The shape memory, which can be one-way or two-way, and superelastic behavior of the material are due to twinning phenomena in the crystal structure of the martensite. One-way shape memory occurs when the alloy is at a temperature where it is martensite and is then subjected to a load sufficient to macroscopically change the shape in a plastic way which also results in a microscopic change to the form of the martensite from twinned to detwinned. When the alloy is heated to the austenite phase, the original shape is recovered. Two-way shape memory is obtained when the part assumes one shape in the martensite phase and another shape in the austenite phase with no applied load. Two-way shape memory behavior is achieved by thermomechanical processing. The superelastic behavior occurs when the alloy is at a temperature where it is austenite and is then subjected to a load which microscopically creates detwinned martensite. When the load is removed, the phase immediately reverts to austenite, since austenite is in equilibrium at the temperature, ~\cite{Kumar2008,Duerig1990} and ~\cite[][ppg. 1035-1048]{Boyer1998}. Above a certain temperature, denoted M$_\text{d}$, superelastic behavior is lost because it is easier for an applied load to move dislocations as opposed to create martensite.

NiTi is appealing for cryogenic applications because one can design parts which exploit the shape memory and/or superelastic behavior around room temperature and then rely on its martensitic material properties at low temperatures. Also of interest to those designing cryostats is that the transformation hysteresis leads to low frequency vibration dissipation in the superelastic state~\cite{Thomson1995}. Complete shape recovery in NiTi diminishes due to the accumulation of defects created by repeated bending, but other shape memory alloys, e.g. ZnCuAu, are more resilient in this regard, \cite{Song2013}.

\subsection{Laminated materials}
\ul{\emph{Flat cable thermal conductivity test specimen:}} We fabricated a specimen to measure the thermal conductivity due to longitudinal heat flow (Figure~\ref{Fig:cable_and_support_structure} (Left)) in a flat cable. The result would allow us to test the assumption that the flat cable thermal conductivity is well predicted by composing the thermal conductivities of its constituent layers (weighted by cross sectional area), whose thermal conductivities have each individually been  measured. The test specimen is a stack of Pyralux coverlay and bondply, and \SI{21.336}{\micron}  [\SI{0.00084}{\inch}] thick Ti 15-3 foil patterned with repeating pattern of straight parallel signal traces, which have been laminated together into flat flexible circuit coupons\footnote{In particular, the coupon stack up is Pyralux LF 7001 (\SI{12.7}{\micron} [\SI{0.0005}{\inch}] Kapton HN and \SI{12.7}{\micron} acrylic adhesive), then \SI{21.336}{\micron} [\SI{0.00084}{\inch}] patterned Ti 15-3, then Pyralux LF 0111 bondply (\SI{25.4}{\micron} [\SI{0.001}{\inch}] acrylic adhesive, \SI{25.4}{\micron} Kapton HN and \SI{25.4}{\micron} acrylic adhesive), then another layer of \SI{21.336}{\micron} Ti 15-3 with the same patterning as the first layer, then a final layer of Pyralux LF 7001.}. We shall refer to this stack composition as stack $A$. The coupons were folded upon themselves and their thermal conductivity was measured in the same way as the Pyralux and Ti 15-3 foil samples, \cite{FilmSampleGeo}, with the patterned straight-line Ti 15-3 traces oriented parallel to the direction of the flow of heat.

\ul{\emph{Flat cable heat transfer coefficient test specimen:}} We fabricated a different test specimen in order to measure the heat transfer coefficient due to transverse heat flow (Figure~\ref{Fig:cable_and_support_structure} (Left)) of a flat cable having very similar constituent material layers. This specimen has three copies of a stack nearly identical to stack $A$, with a thick sheet of copper interleaved between them and adhered to them with acrylic adhesive film. The difference between stack $A$ and the stack used here, which we shall refer to as $A^\prime$, is that the central Pyralux bondply layer is thinner (\SI{12.7}{\micron} [\SI{0.0005}{\inch}] acrylic adhesive, \SI{12.7}{\micron} Kapton HN and \SI{12.7}{\micron} acrylic adhesive) and the Ti 15-3 layers are not patterned. The final ordering and composition of the laminated material stack for use in the thermal boundary resistance test specimen is $CBA^\prime BCBA^\prime BCBA^\prime BC$, where $C$ is oxygen-free electronic copper (`OFE', formerly OFHC oxygen-free high conductivity but same alloy designation 10100) sheet \SI{1.27}{\mm} [\SI{0.05}{\inch}] thick, and $B$ acrylic adhesive film (that which is used in Pyralux) \SI{12.7}{\micron} thick, and it has an overall thickness of \SI{5.740}{\mm} [\SI{\approx 0.226}{\inch}].

Slots were milled into a \SI{73.66 x 12.70}{\mm} [\SI{2.9 x 0.5}{\inch}] bar of the \SI{5.7404}{\mm} [\SI{\approx 0.226}{\inch}] thick material stack to make a serpentine-shaped test specimen, Figure~\ref{Fig:resistance_sample}. This geometry forces heat created on one end of the specimen to flow through 36 copies of the cable stack, $A^\prime$, before it reaches the other end of the test specimen,  a linear distance  of \SI{73.66}{\mm}s away.

\section{Experimental procedure}\label{sec:procedure}
\subsection{Thermal conductivity measurements}
We carried out our thermal conductivity measurements in an Oxford Instruments \SI{75}{\uW} wet helium dilution refrigerator which has a mu-metal cylinder cladding on the outside of its outer vacuum chamber. Based on room temperature readings, we estimate a DC magnetic field strength to be on the order of \SI{<1.6}{\milli \gauss} at the position of the sample, due to the earth.

Figure~\ref{Fig:Sample} (Left) shows an example set up for our thermal conductivity measurements. A wide variety of different mounting geometries were used to accommodate the different sample specimens, which are often determined by the companies that produce them. The following mounting description applies to all specimens, except the heat transfer coefficient specimen. The base of the sample is epoxied to the copper sample mount which has holes that allow it to be bolted to the  copper thermal link to the mixing chamber of the fridge. This thermal link maintains a roughly constant temperature of \SI{40}{\mK} throughout our measurements. The other end of the sample is epoxied to a sample mount that has a heater attachment point, and is otherwise thermally isolated in the fridge except for the manganin heater wiring. Both copper sample mounts are designed to impose a uniform temperature boundary condition. Stycast 1266 epoxy (\cite{Olson1993}) was used for the sample mounted Ti 15-3 (sheet), Ti 21s, NiTi, and Graphite 2020 because low viscosity was desired. High viscosity Hardman 04007 Double Bubble high peel strength epoxy was used for the other samples. The thermometer mounts are machined as narrow (\SIrange{0.762}{1.524}{\mm} [\SIrange{0.030}{0.060}{\inch}]) as possible and are clamped, not glued, into position for the measurement. The narrow width of the mounts serves to minimize the disruption they cause to the temperature gradient along the surface of the sample. The cylindrical temperature sensors (Lake Shore package AA) are coated with the thermally conductive grease Apiezon N, inserted into holes on the thermometer mounts and secured in place with a set screw. To avoid errors due to thermal boundary resistance, the heater and T$_\text{High}$ thermometer are never housed on the same mounting fixture. The same applies to the T$_\text{Low}$ thermometer and the connection of the sample to the fridge. The entire apparatus, including the thermal link to the fridge,  is surrounded by a copper heat shield at the base temperature of the fridge to minimize any radiative heat load.

\begin{figure}[htb]
	\centering
	\includegraphics[width=\textwidth]{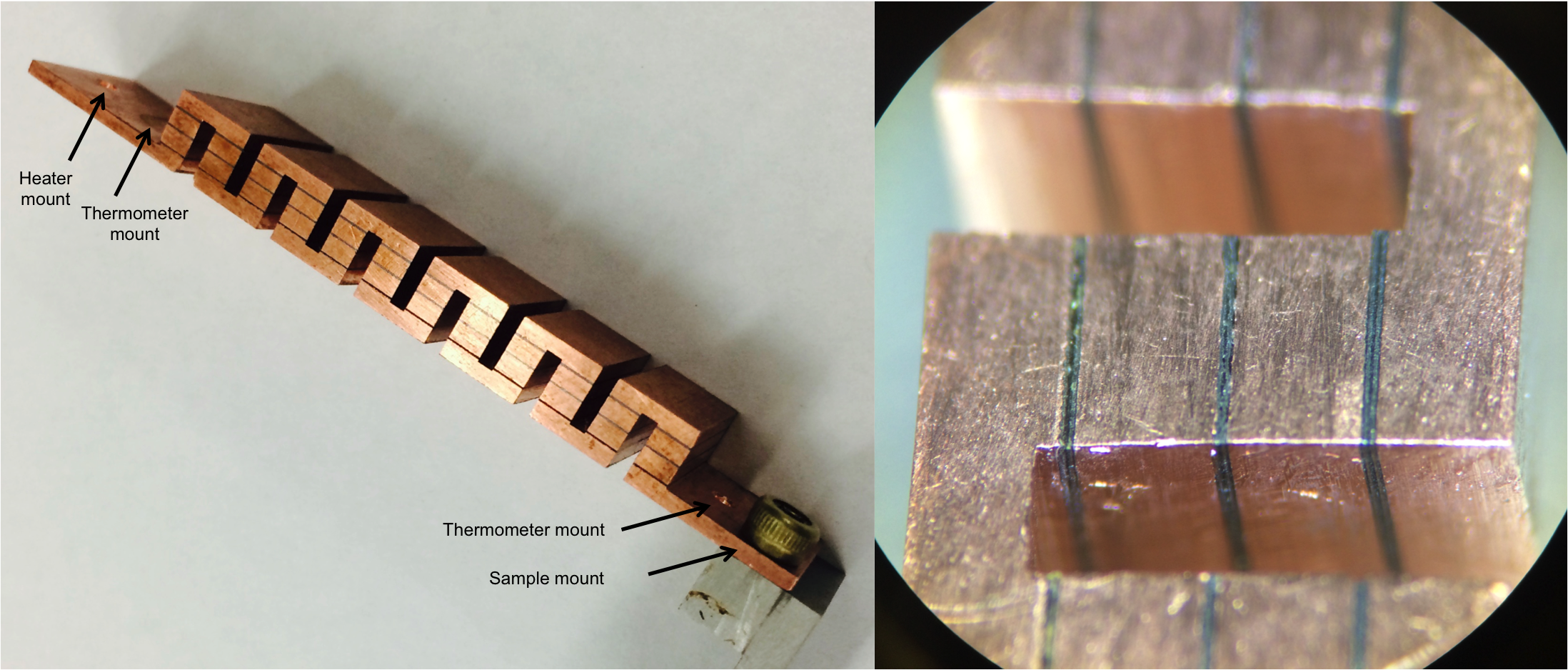}
	\caption{ Left: The serpentine-shaped specimen used to measure the heat transfer coefficient due to transverse heat flow in the flat cable material stack, $A^{\prime}$. The stack, $A^{\prime}$, is laminated with acrylic adhesive (`$B$' from the text) between sheets of copper (`$C$' from the text) in the same way the cable would be attached to copper heat sink points at different fridge temperature stages. Apart from their boundary resistance, the copper sheets do not contribute appreciably to the cable's transfer coefficient because of their very high thermal conductivity. Right: Magnified image of the cable stack-up between the copper layers.}
	\label{Fig:resistance_sample}
\end{figure}

We used a nichrome metal film \SI{20}{\kOhm} resistor for the heater and confirmed its resistance via four-wire measurements at the low temperatures where it was used. The two sample temperatures were measured by germanium temperature sensors and the base temperature was measured by a ruthenium oxide sensor. All temperature sensors and their calibrations were purchased from Lake Shore Cryotronics, Inc. For temperatures below \SI{600}{\milli\K}, the resistances of the  temperature sensors were obtained by DC four wire measurements and transformed into temperatures using the  calibration data. The direction of the DC current in this resistance measurement was periodically flipped to correct for the thermoelectric effect. For higher temperatures, an LR-700 AC resistance bridge was used to read out the temperature sensors. To prevent unwanted heat flow through the wiring, all electrical connections to the sample were connected to the thermal link to the fridge (as labeled in Figure~\ref{Fig:Sample} (Left)) with \SI{76.2}{\micron} [\SI{0.003}{\inch}] diameter manganin wires with lengths greater than \SI{152}{\mm}s [\SI{6}{\inch}es]. This length limited the heat flow through the wires to at most (i.e. maximum over all samples and all temperatures) \SI{2.5}{\%} of the heat dissipated from the heater. The sample was allowed to attain thermal equilibrium before taking each data point. At low temperatures, depending on the material, this step took up to \SI{5}{\hour} per data point. Samples were allowed to fully transition to their superconducting state, should they have one, for the measurements.

From the set of measured thermometer temperatures at different applied powers, we extract the thermal conductivity by relating it to the heat flow through the sample.
\begin{equation} \label{eq:k}
P=\int^{T_{\text{High}}}_{T_{\text{Low}}} \frac{A}{L} K(T) dT
\end{equation}
 A parametric form of $K(T)$ was chosen (see Table~\ref{Tab:Parameter_Fits}), and the parameters of $K(T)$ were estimated by minimizing
\begin{equation}
\chi^2 = \sum_i \left(\frac{P_i - \tilde P_i}{\sigma_{P_i}}\right)^2.
\end{equation}
$\tilde P_i$ and $P_i$ are the powers applied by the heater and expected from equation~\ref{eq:k} for each data point, respectively. $\sigma_{P_i}$ is the estimated error in the power measurement. This method has the benefit of not being sensitive to a large temperature gradient across the sample, but it is restricted by the choice in parameterizing function.

As a check of our parameterization, we calculate the thermal conductivity by assuming $K(T)$ is constant between the two sample thermometer temperatures. The thermal conductivity is well approximated by the following equation as long as $T_\text{High} - T_\text{Low}$ is sufficiently small.
\begin{equation} \label{eq:k_points}
    K\left(\frac{T_\text{High} + T_\text{Low}}{2}\right) = \frac{P\; L}{A(T_\text{High}- T_\text{Low})}
\end{equation}
Equation~\ref{eq:k_points} makes no assumptions about the shape of the overall curve and can capture unexpected temperature dependencies. In particular, we used this method to determine if a power law or logarithmic polynomial function was required to represent the data. The left sides of figures~\ref{Fig:Cable_Mat_K} and~\ref{Fig:Support_Mat_K} show these data points with their associated errors. The more useful parametric functions, which nearly coincide with the constant $K(T)$ values, are plotted on the right side and compared to similar materials measured elsewhere. The difference between the parametric curves and those curves generated assuming constant $K(T)$ are negligible.

We typically measure a small temperature gradient across the sample when the heater is off, suggesting a small, spurious heat load. For our setup, this heat load tends to be less than \SI{5}{\nW} and mostly affects the lowest temperature data points. This effect must be taken into account in the above methods in order to obtain accurate results at the lowest temperatures. Since the base temperature thermometer reading remains almost constant between consecutive temperature data points, we were able to estimate and remove this effect following the work of Didschuns et al.~\cite{Didschuns2004}. Kellaris et al. provide an additional discussion of the analysis methods used in this paper~\cite{Kellaris2014}.

\begin{figure*}[ht]
	\centering
  	\includegraphics[width=\textwidth]{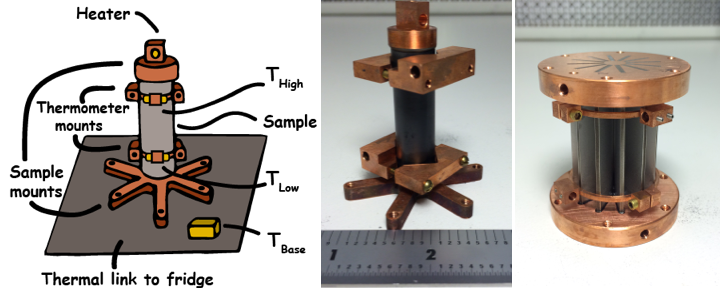}
  	\caption{ Left: A schematic drawing of a typical set up for our thermal conductivity measurement. Center: Our SCP-5050 sample with its thermometer, heater, and fridge mount fixturing.  Right: Our NiTi sample with its thermometer, heater, and fridge mount fixturing. The thermal impedance between the side of the sample bolted to the fridge and the mixing chamber of the fridge is minimal.}
  	\label{Fig:Sample}
\end{figure*}

\subsection{Flat cable heat transfer coefficient measurement} \label{sec:boundary_res}
Because of the discrete layers and thermal boundary resistances (\cite[][are some boundary resistance review articles]{Anderson1981,Swartz1989}) in the flat cable, the thermal conductance due to transverse heat flow does not scale inversely with thickness. Therefore, we cannot define a thermal conductivity for transverse heat flow. Instead we define a `heat transfer coefficient' between the top and bottom face of the flat cable. The heat transfer coefficient, $h$, is equivalent to the aggregate thermal conductivity, $K$, due to transverse heat flow after normalizing by the cable thickness, $t$, i.e. $h=\sfrac{K}{t}$. The measurement technique remains the same as that for the other thermal conductivity measurements in this work.

Forcing heat to flow through 36 copies of the cable stack, $A^\prime$, rather than just one, for example, allows us to establish a measurable temperature gradient. Because the copper thermal conductivity is so high compared to the conductivities of the rest of the materials, the copper layers do not appreciably contribute to the heat transfer coefficient of the system. Since heat transfer coefficients in series add inversely, we multiply our resulting heat transfer coefficient by 36 to obtain the coefficient for just one flex cable. The sample geometry is shown in Figure~\ref{Fig:resistance_sample}.

Referring to Figure~\ref{Fig:cable_and_support_structure} (Left), the heat transfer coefficient allows one to determine, for example, the heat sink contact area needed to effectively thermalize a flat cable running from T$_\text{Hotter}$ to T$_\text{Cold}$. Zobrist et al.~\cite{Zobrist2014} presented a straightforward computational method to calculate this area. The heat transfer coefficient can also be use to calculate the equilibrium temperature of a heat dissipating component attached on one side to a flat cable, which is heat sunk on the other side, exactly as depicted in Figure~\ref{Fig:cable_and_support_structure} (Left). In this case, we know the heat dissipated from the component, $P$, the heat sink temperature $T_\text{Cold}$ and the area of contact between the cable and the component, $A$. The component temperature, $T_\text{Component}$, can be extracted using the equation, 
\begin{equation}
  P = \int^{T_\text{Component}}_{T_\text{Cold}} A \; h(T) dT.
\end{equation}

\subsection{T$_\text{C}$ measurements} \label{sec:Tc}
With the exception of the Nb-47Ti sample, four-wire resistive T$_\text{C}$ measurements were performed in an adiabatic demagnetization refrigerator (ADR) by the collaborators acknowledged. Their measurements were blind in the sense that they were not told the identity of the samples. Each sample was cut to a few \si{ \mm \times \mm} size and then glued with GE varnish to a polished OFE mount with a piece of cigarette paper interposed to prevent shorting. Aluminum-silicon wire bonds were made between the samples and copper traces on a G10 substrate which was also glued to the OFE mount.  The OFE mount was bolted to the coldest stage of the ADR, which does not have active temperature regulation. The T$_\text{C}$s reported are those measured over slow warm-up and correspond to the center value of the superconducting transition. Full transitions occurred over \SI{\sim 1}{\hour} with the most rapid variation in resistance occurring over \SI{\sim 3}{\min}. Resistance readings of the samples and the ruthenium oxide temperature sensor were taken automatically every \SI{\sim 30}{\sec}. The transitions were all easily discernable. The uncertainties quoted are dominated by the uncertainty in the temperature sensor calibration.

For the Nb-47Ti sample, a four-wire resistive T$_\text{C}$ measurement was performed in a BlueFors dry dilution fridge. The sample was glued to an OFE mount and wire bonded to a G10 board in the same way as the other samples. The mount was bolted to the mixing chamber stage of the fridge. The resistance measurement was executed before condensing the mix, while the stage was pre-cooled by a pulse tube cooler. The stage temperature was controlled by a resistive heater and ramped upward at a rate of \SI{5}{\mK \per \minute} while a Lake Shore 370 AC Resistance Bridge recorded both the Nb-47Ti resistance and the temperature from a calibrated ruthenium oxide temperature sensor located next to the mount.

%
%
\section{Results and discussion}\label{sec:results}
%
%

\begin{figure}[t]
  \centering
  \input{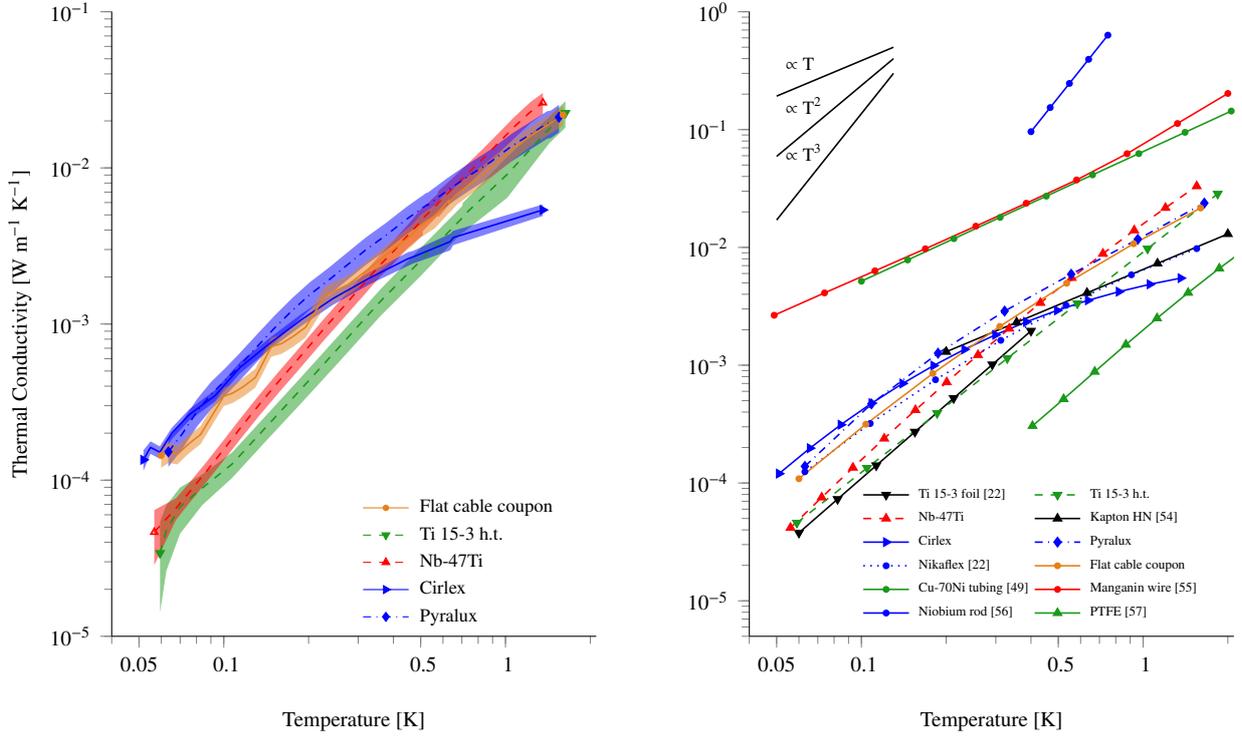}
  \caption{ The thermal conductivities of various materials useful for cryogenic flat flexible cables and wiring are plotted. Left: Materials measured in this work are shown with their thermal conductivities calculated assuming $K(T)$ is constant between sample thermometers. The shaded regions correspond to uncertainty at the \SI{68}{\%} confidence level. Right: Parameter fits to our data are plotted next to thermal conductivity curves found in the literature of other useful cryogenic cabling materials. The data markers do not correspond to measured data points.}
\label{Fig:Cable_Mat_K}
\end{figure}

\begin{figure}[htb]
  \centering
  \input{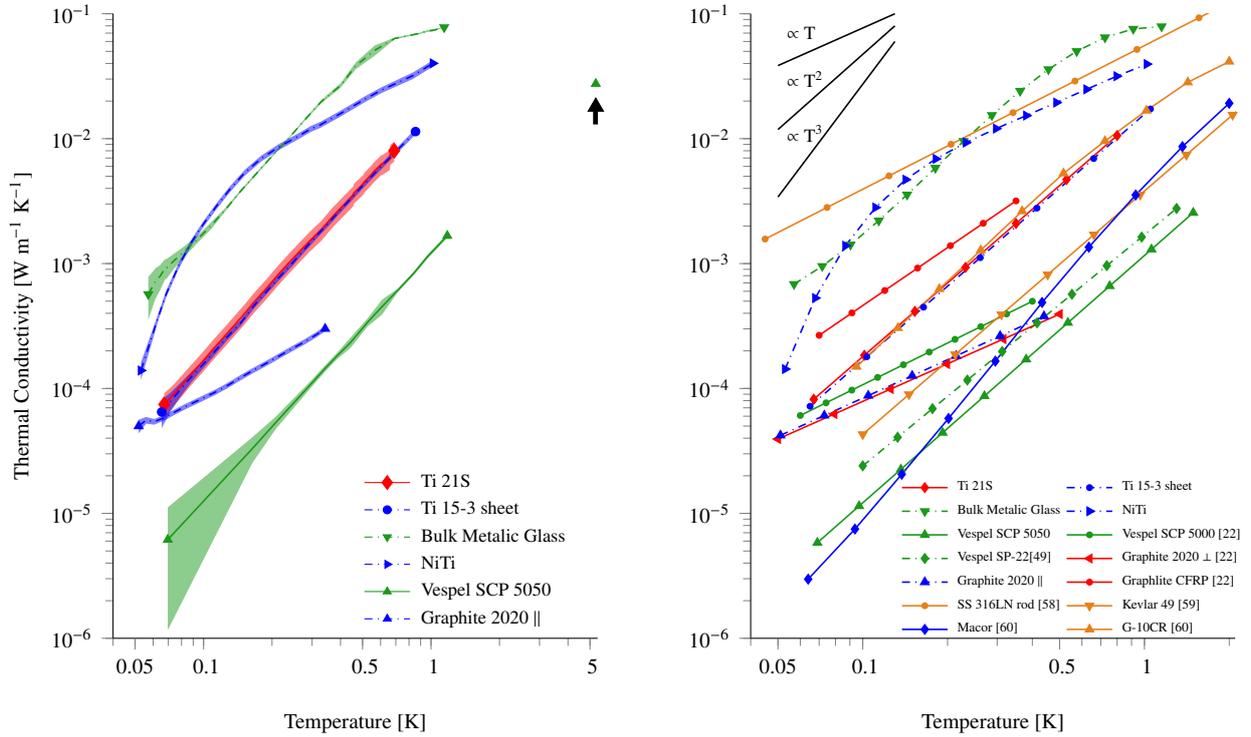}
  \caption{ The thermal conductivities of various materials useful for cryogenic structural supports are plotted. Left: Materials measured in this work are shown 
  with their thermal conductivities calculated assuming $K(T)$ is constant between sample thermometers. The shaded regions correspond to uncertainty at the \SI{68}{\%} confidence level. The arrow points to an additional data point for Vespel SCP-5050, which is \SI[per-mode=fraction,fraction-function = \tfrac]{2.75e-2}{\watt\per\meter\per\kelvin} at \SI{5.3}{\K} . Right: Parameter fits for our data are plotted next to thermal conductivity curves found in the literature of other useful cryogenic support materials. The data markers do not correspond to measured data points.}
\label{Fig:Support_Mat_K}
\end{figure}

\begin{figure}[htb]
  \centering
  \includegraphics[width=\textwidth]{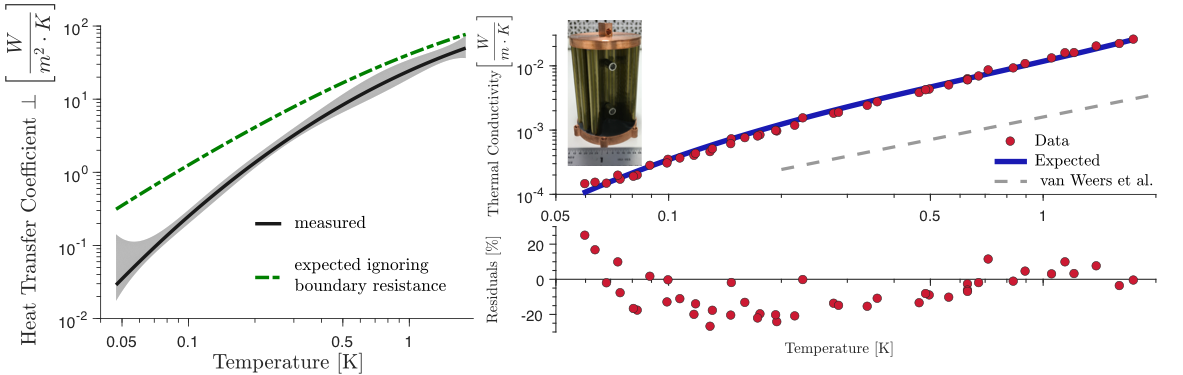}
  \caption{ Left: Heat transfer coefficient, due to transverse heat flow, for an individual flat cable (i.e. one $A^\prime$ stack) derived from the serpentine-shaped heat transfer coefficient test specimen, plotted as a function of temperature. The gray region represents uncertainty at the \SI{68}{\%} confidence level. The result is compared to the expected heat transfer coefficient ignoring any contributions from boundary resistances between layers. Top Right: Data points for our patterned flat cable thermal conductivity (due to longitudinal heat flow) test specimen are plotted on top of the thermal conductivity estimated from composing the conductances of its constituent materials. The fit from the micro-fabricated flat cable of \cite{VanWeers2013} is included for comparison, though the materials and metal patterning are different from our laminated flat cable specimen.  Bottom Right: The difference between the two quantities. Right Inset: A photo of the flat cable text specimen. Vertical metallic traces are apparent.}
\label{Fig:Flex_Cable_Plot}
\end{figure}

Thermal conductivity data points for the  materials discussed in Section~\ref{sec:materials} are shown on the left side of Figures~\ref{Fig:Cable_Mat_K} and~\ref{Fig:Support_Mat_K}, which are grouped to distinguish between cryogenic cabling and support material, respectively. The shaded area of the plots corresponds to the combined measurement and systematic errors at the \SI{68}{\%} confidence level for each material. The error was calculated by considering the uncertainty in the applied sample power, thermometer calibrations, parasitic heat load and sample geometry.

\subsection{Fits to data}
Our thermal conductivity results are compared to those of similar materials measured in the literature on the right side of Figures~\ref{Fig:Cable_Mat_K} and~\ref{Fig:Support_Mat_K}.  Additionally, one can compare the thermal conductivities to that of copper using $K(T) \approx  \frac{\text{RRR}}{0.634}T \si[per-mode=fraction,fraction-function = \tfrac]{\watt\per\meter\per\kelvin}$\footnote{Residual resistivity ratio, RRR, is defined as the quotient of electrical resistivities, $\rho$, measured at, \SI{4}{\K} and \SI{273}{\K}, $\text{RRR} = \sfrac{\rho(\SI{4}{\K})}{\rho(\SI{273}{\K})}$. Typical values for as received copper, 10100 or 10200, range from \numrange{50}{700}, \cite{Fickett1983}.} to $\mathcal{O}(10\%)$ in the vicinity of  $T \sim \SI{1}{\K}$~\cite[][pg. 7-16]{Simon1992}. Our curves are parametric fits to our data $K(T)$ and the parametric functions used are indicated in Table~\ref{Tab:Parameter_Fits}. Power law expressions were used when applicable and may hold outside of the temperature ranges used to fit the curves. For more complicated temperature dependencies, we fit our data to polynomials logarithmic in both $K$ and $T$. These fit types should never be used outside of their specified temperature range because they diverge rapidly from realistic behavior. For NiTi, we used an additional fit coefficient to more accurately represent the measured data. On this table, we also include the total cross-sectional area through which heat flowed, and the distance between the centers of the thermometer mounts for each material measured.

\subsection{Cable component materials}
We find that although the thermal conductivity of the Pyralux LF approaches that of Cirlex at low temperatures, Cirlex is significantly less conductive than Pyralux LF at higher temperatures. We suspect this is due to the acrylic adhesive. The Pyralux LF thermal conductivity can be compared to the Nikaflex CISV coverlay conductivity presented in Kellaris~\cite{Kellaris2014} et al. Nikaflex uses an epoxy adhesive. In terms of availability and manufacturability, the acrylic-, epoxy-, and fluoropolymer-based (which has the lowest dissipation factor and is used in DuPont Pyralux TK laminates) adhesives\footnote{Besides thermal conductivity and dissipation factor, another consideration in choosing a flat cable adhesive product is its bonding temperature. Of the DuPont product family, Pyralux LF bonds at \SIrange{182}{199}{\degreeCelsius}, Pyralux TK bondply at \SIrange{280}{290}{\degreeCelsius}, and Nikaflex CISV at \SI{160}{\degreeCelsius}. At room temperature the CTE  missmatch between the kapton and the titanium alloys are larger than that between kapton and  copper (Table~\ref{Tab:Mechanical_Props}), the industry standard flat cable trace material. The missmatch increases with temperature, which can lead to production yield issues. The problem is exacerbated by the stiffness of the titanium alloys in comparison to copper.} comprise the prevailing adhesive solutions for flexible circuits.

We confirm that mildly heat treating Ti 15-3 foil does not have a measurable effect on the thermal conductivity when comparing our results to those of Kellaris et al.~\cite{Kellaris2014} who measured foil from the same roll that was not heat treated.  The thermal conductivity of the Ti 15-3 foil is slightly less than that of the Nb-47Ti foil sample measured for this paper.

Portions of both the Ti 15-3 and Nb-47Ti foils that we have measured and reported on in this paper have been used to make flat cables for the CDMS experiment\footnote{Fabricated by Basic Electronics Inc., https://www.basicelectronicsinc.com, using Pyralux circuit materials.} to connect between \SIrange{4}{0.015}{\K}. These cables are about \SI{70}{\cm} long, have several trace layers interconnected by plated through holes and have Cirlex stiffened portions where heat sinks, connectors and other surface mount components are attached. At UC Santa Barbara, we have used portions of the Nb-47Ti foil to make microstrip flat cables\footnote{Fabricated by Tech-Etch, https://www.tech-etch.com, using Nikaflex circuit materials.} for the \SIrange{4}{8}{\GHz} range to connect between \SIrange{3.2}{0.1}{\K}, \cite{Walter2017}. Other collaborators at UC Berkeley have used portions of the Nb-47Ti foil to make stripline flat cables operating in the \SIrange{1}{10}{\MHz} range to connect between \SIrange{4}{0.3}{\K}, \cite{Elleflot2018, Avva2018}.

\subsection{Structural support component materials}
The support materials tested show a much broader range in thermal conductivity. Ti 21S and the Ti 15-3 sheet have very similar thermal conductivities. The Ti 15-3 sheet is more conductive than the Ti 15-3 foil. Perhaps this is due to increased impact of phonon surface scattering in the superconducting foil or to changes in microstructure imparted by the thermomechanical processing involved in producing the foil.

We were surprised to measure that the NiTi undergoes a superconducting transition despite its \SI{~50.9}{at\%} nickel content. (Amorphous NiTi with lower nickel content is known to become superconducting, \cite{Lindqvist1992}.) Perhaps the increasing slope of $K(T)$ toward low temperature is a result of the transition. The superconducting transition temperatures we measured (see Table~\ref{Tab:Mechanical_Props}) are similar to previously measured values: \cite[Fig. 1]{Lee1989} suggests \SI{~8.8}{\K} for Nb-47Ti, \cite{Wikus2011} measured \SI[multi-part-units = single]{3.89 \pm 0.01}{\K} for Ti 15-3, and \cite{Rothfuss2011} measured \SI{1.030}{\K} for LM105. Differences in the exact microstructural condition and alloy constitution between the present specimens and those measured previously by other authors might be the source of the T$_\text{C}$ differences.

We present a wider temperature range for the Vespel SCP-5050 in comparison to what has been previously presented~\cite{Kellaris2014}. The two data sets are from specimens of significantly different length, different thermometer mount designs, and different calibrated thermometers. Still, the sets are at worst \SI{5.2}{\%} discrepant within their overlapping temperature measurement range, \SIrange{0.069}{1.49}{\K}, and the error is always larger than the discrepancy between the sets. This is so close that two sets would overlap on Figure~\ref{Fig:Support_Mat_K} so we only plot the new data.

As expected, the similarity between the $\perp$ and $\parallel$ directions of the grade 2020 graphite suggests a high degree of isotropy in the bulk material, perhaps due to use of isotropic filler particles. Both directions of Mersen grade 2020 graphite have thermal conductivity very similar to POCO grade AXM-5Q~\cite{Woodcraft2009b,Kellaris2014}. The thermal conductivity of both these materials are lower than that of AGOT graphite, which is no longer manufactured.

\subsection{Laminated materials}
The flat cable coupon is a layered composite material with (longitudinal) heat flowing parallel to the layers, and each metal trace layer offering a direct, unsevered, path from warm to cool. Under these conditions, an aggregate thermal conductivity can be estimated by summing the heat conductances, i.e. Watts per Kelvin, through each layer as if they were not in contact with each other, and then multiplying the total by the  cable length divided by its cross-sectional area. The thermal conductivity of our flat cable specimen is within \SI{20}{\%} of the value estimated from composing in this manner the heat conductances of its constituent materials throughout the temperature data range, Figure~\ref{Fig:Flex_Cable_Plot} (Right).

Flat cables can also be micro-fabricated by spinning a liquid polyimide precursor into a carrier substrate, depositing metal layers into it via vapor deposition processes, lithographic patterning and etching of traces and finally a soak in solvent to release from the carrier. This process results in down to $\mathcal{O}(\SI{100}{\nm})$ scale thickness flat cables, but is limited in length by the carrier size and is of a different price-scale than the laminated flat cables. As a rough comparison, van Weers et al.\cite{VanWeers2013} presents a thermal conductivity\footnote{The metal fraction of the cross sectional area of our cable and theirs are very different. In the case of our cable, it is about \SI{22}{\%}. However, the heat conduction of dielectrics (including adhesives) used in both cables dominate throughout the temperature range graphed in Figure~\ref{Fig:Flex_Cable_Plot} (Right).} measurement for their micro-fabricated niobium on polyimide cables; they have a factor \num{\sim10} lower thermal conductivity at \SI{1}{\K}. For this approach, material thicknesses can be so small that one has to consider thermal transport dominated by surface scattering.

The heat transfer coefficient due to transverse heat flow was obtained from a measurement of the serpentine-shaped heat transfer coefficient test specimen described in section~\ref{sec:boundary_res}. The left plot in Figure~\ref{Fig:Flex_Cable_Plot} shows the result of our measurement. In contrast to the thermal conductivity due to longitudinal heat flow, the thermal boundary resistance between layers decreases the thermal conductivity from the expected result by a factor of \num{10} at \SI{50}{\mK} and \num{1.5} at \SI{1.8}{\K}. This measurement suggests that special attention may be needed to ensure adequate thermalization for transverse heat flow in laminated constructions. 

\subsection{Caveats}
Finally, we note that microstructural and manufacturing variations occur for all the materials studied in this article. This is to say that material from different manufacturers, or batches, or lots, or at different positions within a single sample, may have different thermal or  mechanical properties. We also note that for most materials our data represents a single sample of the thermal conductivity and T$_\text{C}$ from a single specimen and a single dilution fridge cool down. We have not measured the variations in the materials studied, and we have not measured the effect of repeated thermal cycling in the materials studied.

\sisetup{range-phrase = --}
\ctable[caption={Parameterized thermal conductivity, $K(T)$, for materials measured in this work. The data was fit to a power law if applicable, otherwise a logarithmic polynomial in both $K$ and $T$ was used. The fits assume that $T$ is in units of Kelvin. A is the cross-sectional area of the sample, and L is the center-to-center distance between the sample thermometer mounts. Of the fits which are not power laws, NiTi needs an extra fit coefficient to accurately represent our data. For each material, three significant digits in the fit parameters was determined to be sufficient to reduce any roundoff errors well below our measurement error.},
label=Tab:Parameter_Fits,
width=\textwidth]{>{\raggedright}Xlrrr}{}{
\FL \addlinespace[0.5ex]
Sample & A [\si{\square \centi \meter}] & L [\si{\centi \meter}] & $K(T)$ [\si{\watt\per\meter\per\K}] & Valid Range \ML \addlinespace[1ex]
Pyralux LF & 0.961 & 3.05 & $10^{-1.91 + 1.26\log_{10}(T) + 0.121\log_{10}(T)^2 + 0.353\log_{10}(T)^3}$ & \SIrange{0.063}{1.65}{\K} \NN \addlinespace[1ex]
Cirlex & 1.16 & 2.83 & $10^{-2.33 + 0.544\log_{10}(T) - 0.436\log_{10}(T)^2 + 0.0754\log_{10}(T)^3}$ & \SIrange{0.051}{1.36}{\K} \NN \addlinespace[1ex]
SCP-5050 & 1.30 & 3.11 & $0.00116\;T^{1.98}$ & \SIrange{0.069}{1.49}{\K} \NN \addlinespace[1.5ex]
Graphite 2020 $\parallel$ & 13.0 & 2.35 & $0.000877\; T^{1.02}$ & \SIrange{0.051}{0.44}{\K} \NN \addlinespace[1ex]
LM105 & 0.337 & 3.28 & $10^{-1.11 + 0.252\log_{10}(T) - 2.56\log_{10}(T)^2 - 1.15\log_{10}(T)^3}$ & \SIrange{0.057}{1.15}{\K} \NN \addlinespace[1ex]
Nb-47Ti & 0.871 & 3.05 & $10^{-1.80 + 1.74\log_{10}(T) - 0.292\log_{10}(T)^2 - 0.0270\log_{10}(T)^3}$ & \SIrange{0.056}{1.55}{\K} \NN \addlinespace[1ex]
Ti 15-3 sheet & 1.58 & 2.96 & $0.0157\;T^{1.97}$ & \SIrange{0.065}{1.05}{\K} \NN \addlinespace[1ex]
Ti 15-3 h.t. foil & 0.841 & 3.05 & $0.00911\;T^{1.87}$ & \SIrange{0.059}{1.84}{\K} \NN \addlinespace[1ex]
Ti 21S & 5.38 & 2.67 & $0.0164\;T^{1.96}$ & \SIrange{0.067}{0.80}{\K}	\NN \addlinespace[1ex]
NiTi & 2.12 & 2.96 &  $10^{-1.41 + 0.851\log_{10}(T) -0.808\log_{10}(T)^2 -1.91\log_{10}(T)^3-1.51\log_{10}(T)^4}$ & \SIrange{0.053}{1.02}{\K} \LL \addlinespace[1ex]
}
\sisetup{range-phrase = { to }}

\section{Conclusions}\label{sec:conclusion}
We draw the following conclusions from the information presented.

\begin{enumerate}
\item The $\beta$ phase plus $\alpha$ precipitate titanium alloys Nb-47Ti, Ti 15-3, Ti 21S all have relatively low thermal conductivity at low temperature. Perhaps this is due to their two-phase microstructure. Their high ductility at room temperature, and superconductivity at low temperature make them useful materials for flat cable applications, provided their ductility does not vanish before the cable reaches a stationary position at low temperature. Otherwise, open traces may develop.  With the same low temperature ductility caveat, the high specific strengths of Ti 15-3 and Ti 21S make these materials particularly attractive for use in structural support applications.

\item We find that one can obtain a reasonable estimate of the longitudinal thermal conductivity of a flat cable by composing the previously measured thermal conductivities of its constituent materials. In this case, our thermal conductivity estimates are within \SI{20}{\%} of the actual measured values which, assuming this holds true for other material combinations, is sufficiently accurate to be helpful in the conceptual design of cryogenic flat cables.

\item We find that, as the temperature decreases, the heat transfer coefficient due to transverse heat flow in our flat cable diverges from what one would expect based on the conductivities of its constituent materials. The divergence can be ascribed to the increasing effect of thermal boundary resistances between layers of the cable. This becomes a consideration when one tries to heat sink flat cables and the heat dissipating components that may be attached to them at low temperature. Furthermore, our method of measuring the heat transfer coefficient provides an accurate way to obtain this quantity for different stack combinations.

\item SCP-5050 offers an attractive combination of ductility, stiffness, dimensional stability, and low cryogenic thermal conductivity, arguing for its use in low temperature structural support applications.

\item The 2020 graphite is found to be relatively isotropic in its thermal conductivity. Graphite stands as an inexpensive structural material with relatively low thermal conductivity at low temperature and, unfortunately, low yield strength.  Generally, graphites also have thermal conductivities on par with pure metals around room temperature; they can be used as passive heat switch materials because of this transition from relatively high to relatively thermal low conductivity~\cite{Woodcraft2009b}.

\item Bulk metallic glasses such as LM105 represent a potentially useful low temperature structural support material for their high strength, reported low temperature ductility, and superconductivity. In particular, LM105 has thermal conductivity on par with stainless steel. Especially attractive is the possibility of manufacturing intricate standoff members via an injection molding process.

\end{enumerate}

\section*{Acknowledgements}\label{sec:ack}
This work has been supported in part by the National Science Foundation under the awards PHY-1408597 and PHY-1415388. We thank Curbell Plastics, Liquidmetal Technologies, Fralock and TIMET for providing samples of the SCP-5050, glass phase LM105, Cirlex and Ti 21S, respectively. We thank DuPont, Johnson Matthey Inc., ATI, and Mersen for providing detailed information about their materials. Specifically, we thank Prof. Peter Lee at the National High Magnetic Field Laboratory for the information he provided on NbTi alloys; Phil O’Larey, Technical Fellow at ATI, for working with us to produce the Nb-47Ti foil; Dr. John Fanning at TIMET for the information he provided on titanium alloy metallurgy, Ti 15-3 and Ti 21S; Prof. Betty Young at Santa Clara University, Carl Dawson and Connor FitzGerald for T$_\text{C}$ measurements. Specimens and mounts were machined by the talented mechanicians of the UC Berkeley Physics machine shop.



\clearpage
\section*{References}\label{sec:ref}
\bibliographystyle{elsarticle-num}
\bibliography{./Thermal-Conductivity-Paper-2016}






\end{document}